\documentclass[aps,pre,twocolumn,superscriptaddress,showpacs,floatfix,10]{revtex4}

\usepackage{amsmath}
\usepackage{amssymb}
\usepackage{epsfig}

\newcommand {\dr}{{\mathrm d}\mathbf{r}}

\newcommand {\dd}{{\mathrm d}}
\newcommand {\drr}{{\mathrm d}\mathbf{r}}
\newcommand {\rr}{\mathbf{r}}

\begin{document}

\title{Interfacial and wetting properties of a binary point Yukawa fluid}

\author{Paul\ Hopkins}
 \email[]{Paul.Hopkins@bristol.ac.uk}
  \affiliation{H.H.\ Wills Physics Laboratory, University of Bristol, Tyndall Avenue, Bristol BS8 1TL, UK}

\author{Andrew\ J. Archer}
 \email[]{A.J.Archer@lboro.ac.uk}
  \affiliation{Department of Mathematical Sciences, Loughborough University, Loughborough LE11 3TU, UK}

\author{Robert\ Evans}
 \affiliation{H.H.\ Wills Physics Laboratory, University of Bristol, Tyndall Avenue, Bristol BS8 1TL, UK}

\date{\today}

\begin{abstract}
We investigate the interfacial phase behavior of a binary fluid mixture composed of repulsive point Yukawa particles. Using a simple approximation for the Helmholtz free energy functional, which yields the random phase approximation (RPA) for the pair direct correlation functions, we calculate the equilibrium fluid density profiles of the two species of particles adsorbed at a planar wall. We show that for a particular choice (repulsive exponential) of the wall potentials and the fluid pair-potential parameters, the Euler-Lagrange equations for the equilibrium fluid density profiles may be transformed into a single ordinary differential equation and the profiles obtained by a simple quadrature. For certain other choices of the fluid pair-potential parameters fluid-fluid phase separation of the bulk fluid is observed. We find that when such a mixture is exposed to a planar hard-wall, the fluid exhibits complete wetting on the species 2 poor side of the binodal, i.e. we observe a thick film of fluid rich in species 2 adsorbed at the hard-wall. The thickness of the wetting film grows logarithmically with the concentration difference between the fluid state-point and the binodal and is proportional to the bulk correlation length of the intruding (wetting) fluid phase. However, for state points on the binodal that are further from the critical point, we find there is no thick wetting film. We determine the accompanying line of first-order (pre-wetting) surface phase transitions which separate a thin and thick adsorbed film. We show that for some other choices of repulsive wall potentials the pre-wetting line is still present, but its location and extent in the phase diagram is strongly dependent on the wall-fluid interaction parameters.
\end{abstract}


\maketitle

\section{Introduction}
\label{sec:2cyf_wip_intro}

When charged nano-particles (colloidal macroions) are dispersed in a neutralizing medium the electrostatic potential between the particles is screened by the counterions and is much shorter-ranged than the bare Coulomb interaction between the two identical particles \cite{barrat2003bcs, hansen2000eib}. Derjaguin-Landau-Verwey-Overbeek (DLVO) theory, for example, predicts that the screened electrostatic contribution to the effective interaction potential between the particles takes the form  \cite{hansen2000eib}:
\begin{equation}
\phi(r)=\frac{Q^{*2}}{4 \pi \epsilon_r}\frac{\exp{(-\lambda r)}}{r},
\label{eq:pair_pot_intro}
\end{equation}
where $Q^*$ is the re-normalized charge, $\epsilon_r$ is the dielectric constant of the solvent and $\lambda^{-1}$ is the Debye screening length which determines the thickness of the double layer of opposite charge surrounding each colloidal particle. The DLVO theory also includes a hard-core repulsion, that takes into account the size of the colloids, and dispersion (van der Waals) attraction. In many situations the latter plays a minor role and the effective particle-particle interaction can be modeled by a hard-core, repulsive Yukawa potential. If one considers the limit of high charge and/or a low density of particles, one finds that the particles infrequently come into contact and the effect of the hard-core on the properties of the fluid is small. Under these circumstances one may argue ~\cite{hynninen2003pdh} that Eq.\ \eqref{eq:pair_pot_intro}, applied for all separations $r$, is a good zeroth-order model to describe a charged colloidal suspension. Indeed, Hynninen and Dijkstra~\cite{hynninen2003pdh} showed that the phase diagrams of one-component hard-core repulsive Yukawa particles could be mapped to those of a point Yukawa system for sufficiently large charge on the particles.

The Yukawa potential does, of course, arise in a variety of physical situations. For example, the point Yukawa pair potential may also be used to model the interactions between micron sized dust particles in a charge neutral plasma -- so called dusty plasmas~\cite{piel2002dps}. The binary point Yukawa model has also been used in a recent simulation study of glassy dynamics in low temperature-low density (classical Wigner) glasses~\cite{zaccarelli2008nig}.

In a previous study \cite{hopkins2006pcf}, we investigated the bulk structure and phase behavior of the two component point Yukawa fluid, i.e.\ a binary mixture in which all the potentials between the particles are repulsive Yukawas. We found that for certain choices of the parameters in the pair potentials, and at sufficiently high densities, the mixture separates into two fluid phases, one of which is composed predominantly of particles of species 1 and the other phase is predominantly of species 2. In this paper we investigate the inhomogeneous mixture adsorbed at a single planar wall. Using a simple approximate density functional theory (DFT), we calculate the equilibrium fluid density profiles and thermodynamic quantities relevant for investigating surface phase behavior; the latter are surprisingly rich given the simplicity of the model and DFT.

In Ref.\ \cite{hopkins2006pcf} we showed that by making comparisons with results from the accurate hypernetted-chain (HNC) approximation the simple random phase approximation (RPA) for the pair direct correlation functions $c_{ij}(r)$, with $i,j=1,2$, in the Ornstein-Zernike equation \cite{hansen2006tsl}, yields a reasonably accurate approximation for the bulk fluid radial distribution functions $g_{ij}(r)$ at large and intermediate $r$, provided that the fluid is at a state-point fairly close to the fluid-fluid binodal (specifically, within the Lifshitz line \cite{hopkins2006pcf}). However, the RPA was found to be inaccurate for determining the fluid correlation functions far from the coexistence region, where one must use the more reliable HNC or other related bulk theories \cite{hopkins2006pcf, PhysRevE.47.2676, PhysRevE.54.2827, PhysRevE.57.5988}. We also found that the RPA provides a good approximation for determining thermodynamic quantities such as the bulk fluid Helmholtz free energy. Indeed, the phase diagram resulting from the RPA is in fairly good agreement with that obtained from the HNC~\cite{hopkins2006pcf}.

In the present work we use a simple approximate DFT that, in bulk, generates the RPA for $c_{ij}(r)$ to investigate the inhomogeneous mixture. Based on the experience from studying the bulk fluid mixture \cite{hopkins2006pcf} we expect that the DFT should predict with reasonable accuracy the fluid density profiles and adsorption behavior for state points that are close to the binodal. In such cases, where a fluid close to fluid-fluid phase coexistence is adsorbed at a planar wall, it is possible for there to be wetting of the wall by the coexisting phase, for some choices of the wall-fluid interaction parameters \cite{dietrich12pta}. Specifically if a mixture rich in particles of species 1 is at a state point close to the coexistence line, then we may observe wetting of the wall by a thick film of the coexisting phase that is rich in species 2. This is indeed what we observe, for several different choices of purely repulsive wall potentials. Furthermore, we find that the present system may exhibit a surface phase transition from a thick to a thin adsorbed wetting film and we examine the dependence of this transition on the specific form chosen for the wall potentials.

Some of the inspiration for the present study comes from the particularly elegant DFT studies of wetting pioneered by Sullivan~\cite{sullivan1979vwm} for a one component fluid, where the fluid pair potential featured a hard-core plus an attractive Yukawa tail. The DFT that Sullivan used was similar to the one we use here. The hard-core repulsion (absent in the present system) was treated by making a local density approximation, and the Yukawa attraction treated in a simple mean-field approximation. In a subsequent study of the corresponding binary mixture adsorbed at a hard-wall with exponential wall-fluid attractive potentials, Telo da Gama and Evans~\cite{gama1983aaw} showed that for a particular set of wall-potential and fluid pair-potential parameters~\cite{sullivan1982idp}, the Euler-Lagrange (EL) equations arising from the minimization of the density functional yield a single ordinary differential equation (ODE) that can be integrated directly to determine the equilibrium fluid density profiles. The benefit of this approach is that not only can the density profiles and surface tension be calculated easily without the need for sophisticated numerical schemes, but also that the criteria for different types of film adsorption and locating wetting transitions could be established directly~\cite{sullivan1979vwm,gama1983aaw}. We apply an equivalent method to the two component point Yukawa model, treated within the present DFT, and show, in Appendix A, that for a particular set of fluid-fluid and wall-fluid interaction parameters, we may also derive a single ODE that determines the equilibrium density profiles. However, in the present case the restriction on the parameters precludes fluid-fluid phase separation so that wetting must be investigated by solving the coupled EL equations.

This article proceeds as follows: In Sec.~\ref{sec:2cy_wip_binary_yuk_fluid} we introduce the binary point Yukawa fluid model and recall some key results from Ref.~\cite{hopkins2006pcf}. Sec.~\ref{sec:2cy_wip_inhomo_systems} describes the inhomogeneous fluid situation and defines the external (wall) potentials under consideration. In Sec.~\ref{sec:2cyf_wip_dft} we  briefly describe the approximate DFT that generates the RPA and outline the general DFT approach to calculating adsorption behavior. We present numerical results for the fluid density profiles, adsorption, and for the location of the pre-wetting transition line, for various wall potentials in Sec.~\ref{sec:2cy_wip_results}. Finally in Sec.~\ref{sec:2cyf_wip_discus} we discuss our results and draw some conclusions. 

\section{Background}
\label{sec:2cy_wip_background}

\subsection{The Model Fluid and its Bulk Properties}
\label{sec:2cy_wip_binary_yuk_fluid}

\begin{figure}[tp]
\centering
\includegraphics[width=1.\columnwidth]{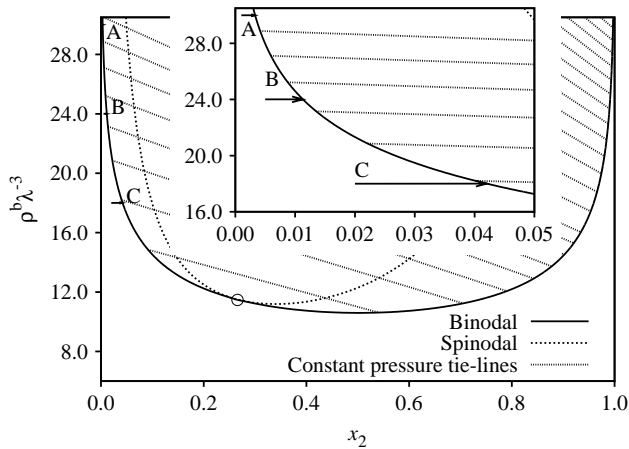}
\caption{\label{fig:phased_paths}
The bulk phase diagram for the binary Yukawa fluid, for the set of pair-potential parameters $M_{11}=1$, $M_{22}=4$ and $\delta=0.1$, calculated using the RPA. $\rho^b$ is the total density and $x_2$ is the concentration of species 2. At sufficiently high densities the fluid demixes. The two-phase region is bounded by the  binodal (solid line), which meets the spinodal (dotted line) at the critical point ($\circ$). The straight tie-lines connect coexisting state-points with pressures $\beta P\lambda^{-3}=150$ to $500$, in increments of $50$, and then from $500$ to $1900$ in increments of $100$ (from bottom to top). The arrows mark three paths which terminate at the binodal (on the species 2 poor side), which are at fixed total densities A: $\rho^b\lambda^{-3}=30$, B: $\rho^b\lambda^{-3}=24$, and C: $\rho^b\lambda^{-3}=18$. The inset shows a magnification of these paths. In subsequent figures we display results for the fluid density profiles at a hard-wall for various $x_2$ along these paths. }
\end{figure}

We consider a binary fluid composed of particles where the pair potentials between particles of species $i$ and $j$ are given by:
\begin{equation}
\phi_{ij}(r)=\frac{M_{ij}\epsilon}{4\pi}\frac{\exp(-\lambda r)}{\lambda r},
\label{eq:yukawa_potential}
\end{equation}
where $\epsilon$ denotes the overall energy scale, $M_{ij}>0$ are the (dimensionless) species specific interaction magnitudes, and $\lambda$ is an inverse decay-length. Following our previous work \cite{hopkins2006pcf}, we allow the inter-species interaction magnitude $M_{12}=(1+\delta)\sqrt{M_{11}M_{22}}$ to depend on a non-ideality parameter $\delta$. For point ions immersed in a medium with inverse screening length $\lambda$, $\delta=0$. However, as discussed in~\cite{hopkins2006pcf}, one can contemplate situations in charged colloidal mixtures where charge renormalization leads to effective potentials with $\delta\neq0$. In the examples that follow, we fix the dimensionless temperature $T^*=k_BT/\epsilon\equiv(\beta\epsilon)^{-1}=1$,  where $k_B$ is Boltzmann's constant, and we set $M_{11}=1$, $M_{22}=4$ and $\delta=0$ or $0.1$. In Ref.~\cite{hopkins2006pcf}, we showed that for an ideal mixture, $\delta=0$, the fluid does not phase separate, while for {\it any} $\delta>0$ there is phase separation at some, sufficiently high, densities. Note that in Eq.\ \eqref{eq:yukawa_potential} there is an additional factor $1/4 \pi$ that was not used in the definition of the pair potentials in Ref.~\cite{hopkins2006pcf}.

The bulk phase diagram for the model fluid with $\delta=0.1$ calculated using the RPA is displayed in Fig.\ \ref{fig:phased_paths}. Details of how this phase diagram is obtained are given in Ref.\ \cite{hopkins2006pcf}. This mixture exhibits a two-phase fluid-fluid coexistence region which is bounded by the binodal (solid line). Within the binodal is the spinodal (dotted line), which is the line in the phase diagram at which the RPA predicts that the compressibility and the bulk correlation length diverge. The binodal and the spinodal meet at a single critical point (circle) in the phase diagram. Within the RPA, scaling the interaction potentials $\phi_{ij}(r)$ by a factor $A$ results in a scaling of the total density, $\rho^b$, by a factor $A^{-1}$. Thus, in Fig.~\ref{fig:phased_paths} the coexisting densities are larger by a factor of $4\pi$ than those in the phase diagram displayed in Fig.~4 of Ref.~\cite{hopkins2006pcf}. Note that the latter figure compares RPA results for the binodal and spinodal with those obtained from the HNC.

The general theory for the asymptotic decay, $r \to \infty$, of the bulk fluid total pair correlation functions $h_{ij}(r)=g_{ij}(r)-1$, states that for a system of particles interacting via short ranged potentials, such as in the present system, $h_{ij}(r)$ may be obtained from the following expression \cite{hopkins2006pcf, evans1994adc}:
\begin{equation}
rh_{ij}(r)=\sum_n A_n^{ij}\exp(i q_n r),
\label{eq:pole_sum}
\end{equation}
where the summation is made over contributions from the poles in the upper half of the complex plane of the function $D(q)$, which is a non-linear combination of the Fourier transforms of the pair direct correlation functions $\hat{c}_{ij}(q)$ \cite{hopkins2006pcf, evans1994adc}. The asymptotic decay, $r \to \infty$, of $h_{ij}(r)$ is determined by the pole $q_n$ with the smallest imaginary part $\alpha_0$ and $A_n^{ij}$ is the amplitude.

In Ref.~\cite{hopkins2006pcf} it was shown that the asymptotic decay, $r \to \infty$, of $h_{ij}(r)$ for the present system, obtained using the RPA, is determined by one of two purely imaginary poles. The location of these poles for the rescaled pair potentials, Eq.~\eqref{eq:yukawa_potential}, is given by:
\begin{equation}
\alpha^{\pm}_{0}=\sqrt{\frac{\rho^b}{2\lambda T^{*}}\left(M_0\pm\sqrt{M_0^2+M_{\delta
}}\right)+\lambda^{2}},
\label{eq:rpasol}
\end{equation}
where $\rho^b$ is the total density, $M_0=(1-x_2)M_{11}+x_2M_{22}$ and $M_{\delta}=4(1-x_2)x_2M_{11}
M_{22}(2+\delta)\delta$~\cite{hopkins2006pcf}. For the case when $\delta=0$, it can be shown that there exists only one pole with $\alpha^+_0>\lambda$. For the case $\delta>0$, there is a second pole, $\alpha^-_0<\lambda$. Of these two poles, the one with the smaller $\alpha_0$ determines the ultimate asymptotic decay of $h_{ij}(r)$ and also defines the bulk fluid correlation length, $\xi \equiv 1/\alpha_0^-$, i.e. $rh_{ij}^{\mathrm{RPA}}(r)\sim A_{ij}^-\exp(-\alpha_0^-r)$, $r\to\infty$. Note that $\xi>\lambda^{-1}$.

\subsection{Wall-Fluid Potentials}
\label{sec:2cy_wip_inhomo_systems}
We investigate adsorption of the model fluid at a planar substrate. A single planar wall exerts the potential $V_i(z)$ on particles of species $i$, where $z$ is the Cartesian axis perpendicular to the wall. We consider here three different wall-fluid potentials. These are all based on the hard-wall, which is infinite for $z<0$ and zero for $z\geq0$:
\begin{equation}
V_i(z)=\begin{cases} \infty & z < 0 \\
0 & z \geq 0, \end{cases}
\label{eq:hw_pot_def}
\end{equation}
for $i=1,2$. Note that this potential is not dependent on any energy or length scale parameters. The hard-wall potential may be modified by adding an extra interaction term for $z\geq0$. This contribution may be repulsive, attractive, or a combination of both. Here we restrict consideration to purely repulsive potentials. Adding an exponential tail maintains the infinite step discontinuity at the origin:
\begin{equation}
V_i(z)=\begin{cases} \infty & z < 0 \\
A_i\epsilon\exp(-\lambda z) & z \geq 0, \end{cases}
\label{eq:exp2_pot_def}
\end{equation}
where the amplitudes $A_i>0$ for both species of particles.
This choice of external potential corresponds to the case where the hard-wall carries a (screened) charge that is uniformly distributed over the surface. Furthermore this potential allows us to recast the EL equations as an ODE~\cite{sullivan1979vwm}.
 We also consider a Yukawa wall potential, similar to the inter-particle interaction potential, of the form
\begin{equation}
V_i(z)=\begin{cases} \infty & z < 0 \\
A_i\epsilon{\exp(-\lambda z)}/{\lambda z} & z \geq 0, \end{cases}
\label{eq:yuk_pot_def_ch6}
\end{equation}
where both amplitudes $A_i>0$. Note that the Yukawa wall potential removes the infinite step discontinuity in $V_i(z)$ at the origin, replacing it by a smooth increase.
%
%
 In both cases we set the wall potential decay length scale equal to the inter-particle interaction length scale $\lambda^{-1}$, in order to simplify the model. This choice therefore precludes any phenomena that may arise from a competition between different length scales in the potentials \cite{archer2002wbg}. We require that the density profiles decay to their bulk values far from the wall,
\begin{equation}
\lim_{z\to\infty}\rho_i(z)=\rho_i^b,
\label{eq:bc_wall}
\end{equation}
for $i=1,2$ and determine the adsorption of species $i$, $\Gamma_i$, which in the present planar geometry is given by the integral
\begin{equation}
\Gamma_i=\int_0^\infty \dd z(\rho_i(z)-\rho_i^b).
\label{eq:gamma_def}
\end{equation}
In regions of the phase diagram where the density profiles vary smoothly as a function of changing state-variable (e.g.\ the concentration or total density), then we also expect $\Gamma_i$ to vary smoothly. Conversely, where the density profiles vary discontinuously as a function of a state-variable, discontinuities will occur in both $\Gamma_1$ and $\Gamma_2$, and will signal surface phase transitions.

\subsection{Implementation of Density Functional Theory}
\label{sec:2cyf_wip_dft}
Here we present a brief description of the DFT approach. For a more complete account see e.g. Refs.~\cite{evans1979nlv,evans1992fif,hansen2006tsl}.  For a fluid composed of $\nu$ different species of particles the thermodynamic grand potential is a functional of  the set of one-body density profiles $\{\rho_i(\rr)\}$, $i=1\dots \nu$:
\begin{equation}
 \Omega[\{\rho_i\}]=F[\{\rho_i\}]-\sum_{i=1}^{\nu}\int\drr\rho_i(\rr)(\mu_i-V_i^{\mathrm{ext}}(\rr)),
\label{eq:omega}
\end{equation}
where $F[\{\rho_i\}]$ is the intrinsic Helmholtz free energy functional, $\mu_i$ is the chemical potential of species $i$ and $V_i^{\mathrm{ext}}(\rr)$ is the external potential acting on species $i$. Minimizing the grand potential functional with respect to variations in the density profiles, one obtains a set of $\nu$ EL equations that may be solved simultaneously to obtain the equilibrium fluid density profiles:
\begin{equation}
\mu_i= \frac{\delta F[\{\rho_i\}]}{\delta\rho_i(\rr)}+V_i^{{\mathrm{ext}}}(\rr).
\label{eq:EL_0}
\end{equation}
The intrinsic Helmholtz free energy functional may be separated into a sum of two contributions: $F[\{\rho_i\}]=F_{\mathrm{id}}[\{\rho_i\}]+F_{\mathrm{ex}}[\{\rho_i\}]$. The first is the ideal-gas term and the second is the excess contribution due to the particle interactions. The ideal gas term is:
\begin{equation}
F_{\mathrm{id}}[\{\rho_i\}] = \sum_{i=1}^{\nu} \beta^{-1}\int\drr\rho_i(\rr)(\ln(\Lambda_i^3\rho_i(\rr))-1),
\end{equation}
where $\Lambda_i$ is the (irrelevant) thermal de Broglie wavelength of particles from species $i$. In the present study we consider a binary mixture ($\nu=2$), and employ a simple mean-field approximation for the excess part of the Helmholtz free energy:
\begin{equation}
F_{\mathrm{ex}}[\{\rho_i\}] = \frac{1}{2}\sum_{i,j=1}^2\int\int\drr\drr'\rho_i(\rr)\rho_j(\rr')\phi_{ij}(|\rr-\rr'|).
\label{eq:rpa_func}
\end{equation}
This functional generates the RPA approximation for the pair direct correlation functions: $c_{ij}(|\rr-\rr'|)=-\beta\frac{\delta^2 F_\mathrm{ex}[\{\rho_i\}]}{\delta\rho_i(\rr)\delta\rho_j(\rr')}=-\beta\phi_{ij}(|\rr-\rr'|)$ \cite{likos2001eis, archer2001bgc}.

Since we only consider external potentials that vary in one Cartesian direction $z$, Eqs.~(\ref{eq:hw_pot_def}--\ref{eq:yuk_pot_def_ch6}), the fluid density profiles also only vary in the $z$-direction. Using Eq.\ \eqref{eq:rpa_func}, we can write Eq.\ \eqref{eq:EL_0} as follows 
\begin{equation}
\mu_i-V_i(z)=\mu_{\mathrm{id},i}(z)+\sum_{j=1}^2\int\dr' \phi_{ij}(|\rr-\rr'|)\rho_j(z'),
\label{eq:EL_def_1}
\end{equation}
where $\mu_{\mathrm{id},i}(z)=\beta^{-1}\ln(\Lambda_i^3\rho_i(z))$ is the ideal-gas contribution to the chemical potential of species $i$. Given the external potentials and the boundary conditions in Eq.\ \eqref{eq:bc_wall}, we may solve the EL equations \eqref{eq:EL_def_1} to obtain the equilibrium fluid density profiles. Generally, the solutions must be obtained numerically, for example, by using a simple Picard iterative scheme. However, for a particular choice of (exponential) wall and particle interaction parameters, one can show that the EL equations may be transformed to yield a single ODE that can be integrated once to give an explicit quadrature for the fluid density profiles. The derivation is outlined in Appendix A.

\section{Results of Calculations}
\label{sec:2cy_wip_results}

We have performed calculations for a variety of bulk state points and for various choices of the wall-fluid and inter-particle potential parameters. Representative results are presented here.

\subsection{Exponential wall and $\delta=0.0$}
\label{sec:res_anal_mod}
\begin{figure}[t]
\centering
\includegraphics[width=1.\columnwidth]{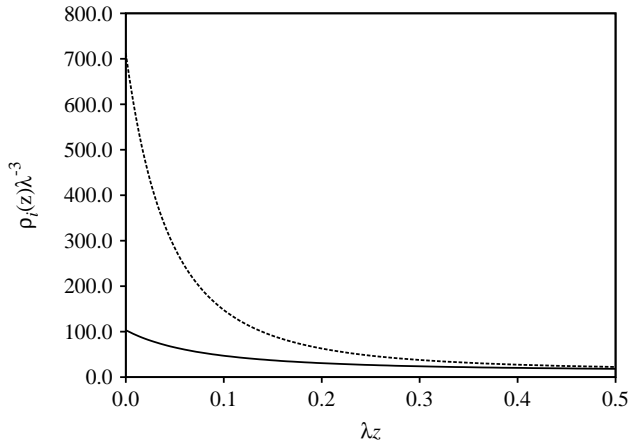}
\caption{\label{fig:sullivan}
The density profiles $\rho_i(z)$ of a binary point Yukawa fluid at an exponential wall, Eq.~\eqref{eq:exp2_pot_def}, calculated by solving the ODE, i.e. from Eq.~\eqref{eq:sullivan_8} and \eqref{eq:sullivan_3}. The solid line is $\rho_1(z)$ and the dashed line is $\rho_2(z)$. The parameters for the fluid pair potentials are $M_{11}=1$, $M_{22}=4$, and $\delta=0$, and the repulsive exponential wall parameters are $A_1=1$ and $A_2=\sqrt{M_{22}/M_{11}}A_1=2$. The bulk fluid at $z \to \infty$ has $\rho^b\lambda^{-3}=30$ and $x_2=0.5$. The density profiles exhibit strong adsorption at the wall but decay rapidly and monotonically to the bulk values.
}
\end{figure}

For the particular case when $\delta=0$, and the exponential wall potential~\eqref{eq:exp2_pot_def} with $A_2=\sqrt{\frac{M_{22}}{M_{11}}}A_1$, we can use the approach described in Appendix \ref{sec:sullivan} to calculate the fluid density profiles. Recall that for $\delta=0$ the fluid does not demix at any density \cite{hopkins2006pcf}. From the analysis of Appendix A we find that the density profiles, $\rho_i(z)$ are both monotonic functions of $z$, and the densities at the wall are determined by Eq.~\eqref{eq:sullivan_10}. For the given set of pair potential parameters, the wall-fluid potential magnitude, $A_1$, determines the behavior of the fluid at the wall. For $A_1\to0$, the wall-fluid potential is equivalent to a hard-wall and we find strong adsorption of both species at the wall. Recall that for a hard wall, the sum of the contact densities is equal to the bulk fluid pressure divided by $k_BT$ and the pressure is large for these mixtures. For $A_1\sim1$ there is again strong adsorption at the wall, but the value of the contact density is decreased. For $A_1\gg1$ then $\rho_1(0)\propto e^{-\beta \epsilon A_1}$ resulting in a depletion of both species at the wall and monotonically increasing profiles. For all state points each density profile returns to its bulk value over a distance $\sim\xi$, the bulk correlation length. In Fig.\ \ref{fig:sullivan} we display a typical set of density profiles showing such behavior for the case when the total bulk density $\rho^b\equiv \rho_1^b+\rho_2^b=30\lambda^3$ and the bulk concentration $x_2 \equiv \rho_2^b/\rho^b=0.5$.

In order to understand the adsorption at the wall it is important to consider the contributions to the free energy that arise from the fluid-fluid and fluid-wall interactions. We first consider the simplest case, $A_1\to0$. The introduction of the hard-wall results in the total absence of any fluid behind the wall, i.e. $\rho_i(z<0)=0$, for $i=1,2$. Therefore, any particles close to the wall have reduced their fluid-fluid interaction energy, which in turn favors strong adsorption of both species at the wall. Since particles of species 2 have a larger (repulsive) interaction energy, the decrease in the free energy is greater for species 2, which results in stronger adsorption of species 2 at the wall. However, as $A_1$ increases there is an additional free energy contribution for particles close to the wall due to the increased wall-fluid repulsive interaction. If $A_1$ is increased sufficiently then this energy contribution negates the benefit from particles adsorbing at the wall and the adsorption is reduced. As $A_1$ is increased still further then ultimately both species become depleted at the wall, since the wall-fluid interaction becomes the dominant contribution to the free energy. This rather simple behavior is not unexpected. Recall that the binary fluid does not demix for $\delta=0$, so we do not observe any wetting or layering behavior.

\subsection{Hard wall and $\delta=0.1$}
\label{sec:res_hardwall}

\begin{figure}[t]
\centering
\includegraphics[width=1.\columnwidth]{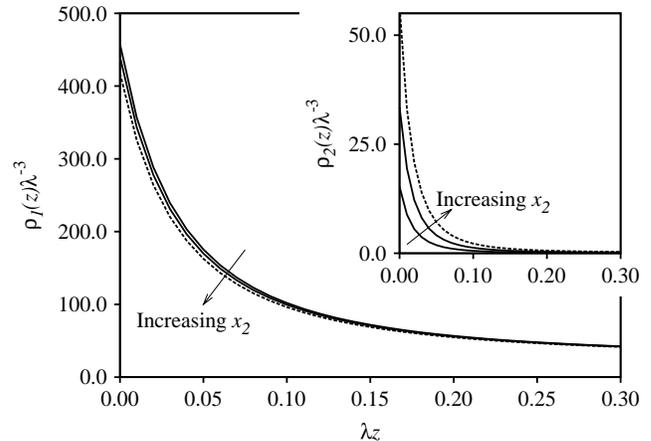}
\caption{\label{fig:pathA}
The density profiles $\rho_i(z)$ for the point Yukawa mixture with $\delta=0.1$ at a hard-wall, Eq.~\eqref{eq:hw_pot_def}, calculated for the fluid at state points along path A in Fig.~\ref{fig:phased_paths}. The density profiles of species 1 are shown in the main figure and the species 2 profiles are shown in the inset. For all concentrations $x_2$, up to and including bulk coexistence, both density profiles show strong adsorption at the wall, but decay to the bulk densities over a distance $\sim \xi$, the bulk fluid correlation length. Results are shown for $x_2=0.0011$, $x_2=0.0022$ and $x_2=0.0033\simeq x_{2,\mathrm{coex}}$, the coexistence concentration for this total density, $\rho^b\lambda^{-3}=30$. The density profiles for the fluid at coexistence are those with the dashed lines.}
\end{figure}

\begin{figure}[t]
\centering
\includegraphics[width=1.\columnwidth]{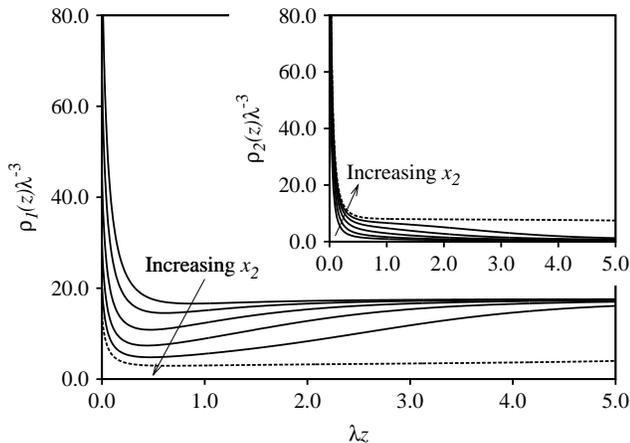}
\caption{\label{fig:rho_dens_hw_C}
As in Fig.~\ref{fig:pathA}, except here the density profiles are calculated along path C in Fig.\ \ref{fig:phased_paths} at constant total density $\rho^b\lambda^{-3}=18$. For concentrations close to the binodal the density profiles exhibit a thick wetting film rich in species 2 intruding between the bulk fluid and the wall. The thickness of this film diverges logarithmically as $x_2 \to x_{2,\mathrm{coex}}$, where $x_{2,\mathrm{coex}}\simeq0.04207$ is the bulk coexistence value. The density profiles displayed are for concentrations from $x_2=0.022$ to $x_2=0.042\simeq x_{2,\mathrm{coex}}$, in increments of $0.004$. The dashed-line density profiles are those for $x_2\simeq x_{2,\mathrm{coex}}$ where the wall is almost completely wet by the phase rich is species 2.}
\end{figure}

For mixtures with $\delta=0.1$ the coupled integral equations~\eqref{eq:EL_def_1} must be solved numerically, rather than using the approach described in Appendix A. In this sub-section we describe the behavior of the fluid at the hard-wall, with potentials given by Eq.\ \eqref{eq:hw_pot_def}. We find that the adsorbed fluid exhibits different types of behavior, depending on the state-point. However, one common feature is that there is always strong adsorption at the wall. The contact densities, $\rho_i(0)$, for both species of particles can be rather large, and this effect is particularly marked in the case of the particles of species 2, which are always favored by the hard-wall, and stems from the high pressures in these systems. For state-points far away from bulk coexistence we find that for both species the density profiles exhibit a thin adsorbed layer at the wall, and decay to their bulk values over a distance $\sim \xi=1/\alpha_0^->\lambda^{-1}$. For all state-points on the right side of the binodal in Fig.~\ref{fig:phased_paths}, i.e. rich in species 2, this type of decay persists up to and including the binodal. However, to the left of the binodal in Fig.~\ref{fig:phased_paths}, i.e.\ for state-points poor in species 2, we find that in some instances the adsorbed layer grows much thicker on approaching the binodal, while for some other state points only a thin adsorbed layer remains. In order to illustrate this, we display results for density profiles calculated along three different paths, increasing $x_2$ towards the binodal, at constant total densities. These three paths A, B, and C, at successively lower densities, are displayed in Fig.~\ref{fig:phased_paths}.

On path A, with total density $\rho^b\lambda^{-3}=30$, we find that for all concentrations, the density profiles decay to their bulk values over a finite distance $\sim\xi$, so that the adsorptions, $\Gamma_1$ and $\Gamma_2$, remain finite up to and including the coexistence state point. A series of density profiles  along path A are displayed in Fig.~\ref{fig:pathA}. Note the strong adsorption and high contact densities at the wall. Along path C, at constant density $\rho^b\lambda^{-3}=18$, we find that the density profiles are radically different for state-points close to coexistence. Instead of there being a thin (finite) adsorbed layer at the wall, we find a thick film of fluid rich in species 2 adsorbed at the wall. As $x_2 \to  x_{2,\mathrm{coex}}$, where $ x_{2,\mathrm{coex}}$ is the concentration at the binodal, the thickness of the wetting film increases and ultimately diverges, i.e. the adsorption $\Gamma_2\to+\infty$ as $x_2\to x_{2,\mathrm{coex}}$. This behavior is termed complete wetting. A number of density profiles along path C are displayed in Fig.\ \ref{fig:rho_dens_hw_C}, indicating the growth of the wetting film as $x_2 \to x_{2,\mathrm{coex}}$. Note that the density profiles of species 1, $\rho_i(z)$, are non-monotonic. As $x_2$ increases, the contact density $\rho_1(0)$ reduces, and a minimum develops in the profile at $\lambda z \lesssim0.5$. The growth of the wetting film is accompanied by increasing depletion of species 1, and $\Gamma_1\to-\infty$ as $x_2\to x_{2,\mathrm{coex}}$.

Recall that for a one-component fluid, exhibiting liquid-gas phase coexistence, one observes that for a bulk gas with density near to coexistence, a thick wetting film of the liquid may be adsorbed at the planar substrate or wall. In the case when all the potentials are short ranged, it can be shown that the thickness of this film, $l$, diverges as $l \sim -l_0\ln|\rho_g-\rho^b|$ where $\rho_g$ is the density of the gas at coexistence. The amplitude $l_0$ depends on the relative ranges of the wall-fluid and the fluid-fluid potentials \cite{dietrich12pta,archer2002wbg}. Provided $l$ is large, the adsorption at the wall $\Gamma$ is proportional to $l$: $\Gamma\simeq(\rho_l-\rho_g)l$, where $\rho_l$ is the density of the coexisting liquid. In an entirely analogous way, in the present two component system, as $x_2$ approaches coexistence at constant $\rho^b$, the adsorption of species 2, $\Gamma_2$, is given by
\begin{equation}
 \Gamma_2\sim-l_0(\rho_2^{b,\beta}-\rho_2^{b,\alpha}) \ln|x_2-x_{2\mathrm {,coex}}|,
\label{eq:gamma_correl}
\end{equation}
where $\rho_2^{b,\beta}$ is the bulk coexisting density of the (wetting) phase $\beta$ rich in species 2 and $\rho_2^{b,\alpha}$ is the same quantity in phase $\alpha$, poor in species 2 ~\cite{archer2002wbg}. For the present wall potentials, Eqs.~\eqref{eq:hw_pot_def}--\eqref{eq:yuk_pot_def_ch6}, we expect the length scale $l_0=\xi_w$, the bulk correlation length of the fluid phase rich in species 2 that wets the wall. The adsorption calculated from the DFT via Eq.~\eqref{eq:gamma_def} is plotted for state-points along path C in Fig.~\ref{fig:gamma_resultsB}a where we also compare with the asymptotic result in Eq.~\eqref{eq:gamma_correl} with $l_0=\xi_w=1/\alpha_0^-$, where $\alpha_0^-$ is the imaginary part of the pole that determines the asymptotic decay of $h_{ij}(r)$ in the wetting phase, obtained from Eq.\ \eqref{eq:rpasol}. Note that since $\alpha_0^-$ depends on the total bulk density, it is different in the two coexisting phases. The asymptotic formula for $\Gamma_1$, equivalent to Eq.~\eqref{eq:gamma_correl}, provides an equally good fit to the DFT results for this quantity.

It is important at this stage to understand why we get wetting for $\delta=0.1$, when for $\delta=0$ we did not. For $\delta=0.1$ there is an added energy cost for particles of species 1 and 2 to mix, which in bulk drives phase separation at sufficiently high densities. It is the combination of this effect and the tendency of the fluid to adsorb at the wall that induces wetting. The high densities of both species at the wall promotes local phase separation; in this case creating a region rich in species 2 and poor in species 1 which increases as the state-point approaches the binodal.

\begin{figure}[t]
\centering
\includegraphics[width=1.\columnwidth]{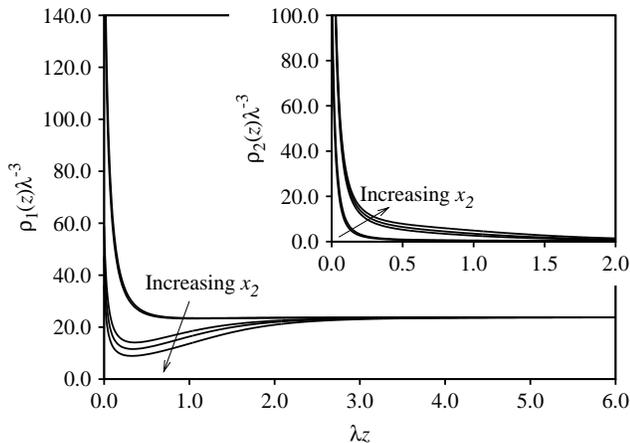}
\caption{\label{fig:rho_dens_hw_B}
As in Fig.~\ref{fig:pathA}, except here the density profiles are calculated along path B in Fig.\ \ref{fig:phased_paths} at constant total density $\rho^b\lambda^{-3}=24.0$. The concentrations are $x_2=0.0091,$ 0.0095, 0.0096, 0.0097 and 0.0098. We observe a discontinuous change in the density profiles as $x_2$ is changed continuously. The thin to thick adsorbed film transition (pre-wetting) occurs between $x_2=0.0095$ and 0.0096. The coexistence concentration at this total density is $x_{2,\mathrm{coex}}=0.0113$.}
\end{figure}

We now consider the fluid interfacial behavior along path B, which has constant total density $\rho^b\lambda^{-3}=24.0$. As we increase $x_2$, we find initially that the density profiles decay rapidly to the bulk values over a short distance $\sim \xi$, similar to the density profiles on path A. However, as we increase the concentration $x_2$ further, we find a discontinuous change in the density profiles which leads to a discontinuous change in the adsorptions $\Gamma_i$. Increasing $x_2$  towards the coexistence value, we find that the profiles are similar to those on path C and that $\Gamma_2$ diverges in a manner similar to that described for path C as $x_2\to x_{2\mathrm{,coex}}$. This discontinuous change in the adsorption denotes a point on the pre-wetting phase transition line, which is a line in the phase diagram that separates regions with thick and thin adsorbed films~\cite{cahn1977cpw}. In Fig.~\ref{fig:rho_dens_hw_B} we display a number of density profiles calculated on path B, showing the discontinuous change on crossing the pre-wetting line. In Fig.~\ref{fig:gamma_resultsB}b we display the DFT results for $\Gamma_2$, calculated from the density profiles, and compare with the asymptotic result in Eq.~\eqref{eq:gamma_correl}, where once again $l_0=\xi_w=1/\alpha_0^-$ is the bulk correlation length of the wetting phase, calculated from Eq.~\eqref{eq:rpasol}.

\begin{figure}[t]
\centering
\includegraphics[width=1.\columnwidth]{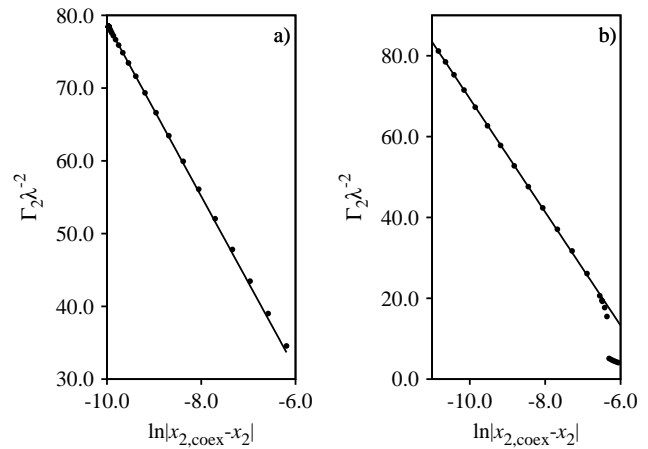}
\caption[Plot of $\Gamma_2$ against $\ln|x_2-x_{2,\mathrm{coex}}|$ along path B.]{\label{fig:gamma_resultsB}
The adsorption of species 2, $\Gamma_2$, (points) at a hard-wall plotted versus $\ln|x_2-x_{2,\mathrm{coex}}|$, the logarithm of the difference between the species 2 concentration and the value at bulk coexistence, calculated a) along path C in Fig.~\ref{fig:phased_paths}, which is at total density $\rho^b\lambda^{-3}=18.0$, and b) along path B, where $\rho^b\lambda^{-3}=24.0$. The solid line is the asymptotic result in Eq.\ \eqref{eq:gamma_correl}, where $l_0=\xi_w$ is the bulk correlation length in the phase wetting the wall, calculated from the RPA -- see text. In b) we observe a jump in $\Gamma_2$ at $\ln|x_2-x_{2,\mathrm{coex}}|\simeq-6.3$, which corresponds to the pre-wetting transition.}
\end{figure}

\begin{figure}[tp]
\centering
\includegraphics[width=1.\columnwidth]{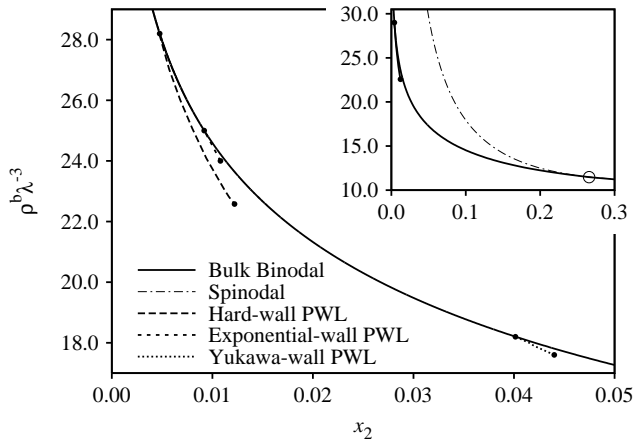}
\caption{\label{fig:exp_yuk_hw_pwl}
The pre-wetting lines (PWL) for the binary point Yukawa fluid at a hard-wall and at hard-walls augmented by repulsive exponential and Yukawa tails. The solid line is a portion of the bulk binodal from Fig.~\ref{fig:phased_paths}. The main figure is a magnification of the inset, which displays the spinodal, the bulk critical point (circle) and the hard-wall PWL. All the PWLs descend tangentially from the binodal and terminate in a critical point, denoted by the symbol $\bullet$. The hard-wall PWL meets the binodal at $\rho^b\lambda^{-3}=28.2$ and the critical point is located at the density $\rho^b_{\mathrm{ c}}\lambda^{-3}=22.5$ and concentration $x_{2,\mathrm{ c}}=0.012$. The PWL for the hard-wall with repulsive exponential tail Eq.~\eqref{eq:exp2_pot_def} is for the case with parameters $A_1=1.0$, $A_2=2.0$. Much further down the binodal is the PWL for the hard-wall with repulsive Yukawa tail, Eq.~\eqref{eq:yuk_pot_def_ch6}, with parameters $A_1=1.0$, $A_2=2.0$. The PWLs for the exponential and Yukawa walls are much shorter than the hard-wall PWL, although by varying $A_1$ and $A_2$ the location and extent of the PWL may be varied -- see text.}
\end{figure}

In order to establish the location of the pre-wetting transition line, it is necessary to calculate the grand potential, $\Omega$. In the vicinity of the pre-wetting transition we find that there are two branches of solutions which minimize the grand potential -- one branch corresponding to a thin adsorbed film and the other to a thick adsorbed film. We compute the value of $\Omega$ for each set of profiles to establish which set is the global minimum of the grand potential. In practice, we calculate $\Omega[\{\rho_i\}]$, scanning along lines of constant total density in the phase-diagram, for both increasing and decreasing concentration, $x_2$. By plotting $\Omega$ versus $x_2$ we determine the concentration where these two branches intersect (where the grand potentials of the thick and thin film branches are equal) which is the pre-wetting concentration at the particular total density. Since the gradient of the minimum grand potential changes discontinuously and the adsorptions, $\Gamma_i$, jump discontinuously, this is a first order phase transition. By repeating this procedure for different total densities, we are able to map out the location of the pre-wetting line in the phase diagram. We find that it extends tangentially from the binodal, as discussed in Ref.~\cite{hauge1983caf}, and terminates in a pre-wetting critical point, where the difference in adsorption $\Gamma_i$ between the thin and thick adsorbed film vanishes. The hard-wall pre-wetting line (long dashed line) is displayed in Fig.~\ref{fig:exp_yuk_hw_pwl}. As the inset shows, the line lies rather close to the bulk binodal and extends over a narrow range of densities.

\subsection{Repulsive Exponential and Yukawa walls}
\label{sec:rep_exp_yuk_wall}
Since the hard-wall potentials do not have any adjustable parameters, the location of the pre-wetting line is fixed for a given set of fluid-fluid interaction parameters. However, if we consider a hard-wall augmented by repulsive exponential or Yukawa tails, Eqs.~\eqref{eq:exp2_pot_def} and \eqref{eq:yuk_pot_def_ch6}, this introduces two independent parameters (the amplitudes $A_i$) which may be varied. Studying various combinations of these parameters, we find that the occurrence of wetting, and the existence and location of a pre-wetting line is strongly dependent on the magnitudes of $A_1$ and $A_2$. For $A_1=A_2\to0$ we recover the hard-wall behavior. For combinations of parameters where $A_i\simeq \sqrt{M_{ii}}$ with $i=1,2$, we find that with both the exponential and Yukawa walls, there is a pre-wetting line that is not drastically removed in the phase diagram from the location of the hard-wall pre-wetting line -- see Fig.~\ref{fig:exp_yuk_hw_pwl} for examples with $A_1=1$ and $A_2=2$. If we increase the ratio $A_2/A_1$, we find that the pre-wetting line moves down the binodal, towards the bulk critical point. As it approaches the critical point, there are indications that the wetting transition may change from a first order to a continuous (critical) wetting transition, although we have not investigated this issue in detail -- when locating the wetting transition numerically, it can be difficult to discriminate between a first order wetting transition with a very short pre-wetting line and a true continuous wetting transition. If $A_2/A_1$ is increased even further the wetting transition and all wetting behavior may be pushed onto the opposite branch of the binodal, i.e. any wetting is by the phase rich in species 1. Conversely, if the ratio $A_2/A_1$ is decreased we find that the pre-wetting line moves up the binodal creating a larger complete wetting regime. In all cases where there is complete wetting, we confirmed that the increase of $\Gamma_2$ as the binodal is approached is given by Eq.~\eqref{eq:gamma_correl} with $l_0=\xi_w$, irrespective of the magnitudes of $A_1$ and $A_2$.

\section{Concluding Remarks}
\label{sec:2cyf_wip_discus}
Using a simple mean field DFT, we have investigated the interfacial behavior of a two component point Yukawa fluid adsorbed at a planar wall. For a restricted set of the parameters in the model, the EL equations reduce to a single ODE that can be integrated to yield a simple expression, Eq.~\eqref{eq:sullivan_8}, for the density profiles. However, the restriction on the parameters required to make this simplification limits the method to cases where the fluid does not exhibit fluid-fluid phase separation, i.e. $\delta=0$, and therefore precludes the study of wetting behavior. This scenario differs from that in the binary mixture of hard-core plus attractive Yukawa tail particles adsorbed at a hard-wall augmented by exponential attractive tails where the equivalent mixing rules lead to different classes of wetting behavior~\cite{gama1983aaw}. In the general unrestricted case, we must obtain the density profiles numerically by solving either the coupled pair of integral EL equations in Eq.\ \eqref{eq:EL_def_1}, or for the exponential wall potential Eq.~\eqref{eq:exp2_pot_def}, the coupled pair of differential equations in Eq.\ \eqref{eq:sullivan_2}. For the model fluid mixture that exhibits phase separation (where the parameter $\delta>0$), we find generally that particles of species 2 are more strongly adsorbed at the wall than are particles of species 1. Of course, one could choose the wall potential parameters, by making $A_2\gg A_1$ in Eqs.~(\ref{eq:exp2_pot_def}) and (\ref{eq:yuk_pot_def_ch6}), so that the effective attraction between the wall and the particles of species 1 is sufficiently strong that this situation is reversed. In the more common case, where there is a stronger effective attraction between the wall and the particles of species 2, we find that for some state points on the species 1 rich side of the bulk binodal, that a thick wetting film rich in species 2 forms at the wall. In particular this occurs for a hard wall. We find that in the complete wetting regime, the thickness of this wetting film increases logarithmically as the bulk concentration approaches its value at coexistence.
 
Generally we find a first-order wetting transition with a pre-wetting line extending out of bulk coexistence for the various wall potentials that we have considered. However, the location of this pre-wetting line is very sensitive to the precise form of the wall potentials and to the values of the parameters in the various potentials. It is important to note that for all the wall potentials that we considered, the decay length of the potentials is $\lambda^{-1}$, which is the length scale in all the fluid pair-potentials -- see Eq.\ \eqref{eq:yukawa_potential}. We have not investigated the influence on the wetting behavior arising from varying this length scale. We expect such a modification to have a strong influence, not only on the existence and location of the pre-wetting transition line, but also on the `rate' of the growth of the thick wetting film [c.f.\ Eq.~\eqref{eq:gamma_correl}]. In a study of a different binary fluid, with short ranged Gaussian potentials, exhibiting bulk fluid phase behavior that is similar to the present system \cite{archer2001bgc, archer2002wbg}, it was found that if the wall potentials are of the form in Eqs.\ \eqref{eq:exp2_pot_def} or \eqref{eq:yuk_pot_def_ch6}, but with with a modified decay length $\gamma^{-1}$ (i.e.\ with the length $\lambda^{-1}$ replaced by the decay length $\gamma^{-1}$ in these potentials), then the amplitude $l_0$ in Eq.~\eqref{eq:gamma_correl} is no longer necessarily the bulk correlation length of the wetting phase, $\xi_w$. In the case of the exponential wall-potential Eq.~\eqref{eq:exp2_pot_def}, the longer of $\gamma^{-1}$ and $\xi_w$ determines the prefactor $l_0$ in Eq.\ \eqref{eq:gamma_correl} \cite{archer2002wbg, dietrich12pta}. In the case of the Yukawa wall-potential \eqref{eq:yuk_pot_def_ch6}, if $\xi_w>\gamma^{-1}$, then $l_0=\xi_w$. However, when $\xi_w<\gamma^{-1}$, then the prefactor $l_0$ in Eq.\ \eqref{eq:gamma_correl} is neither $\xi_w$ nor $\gamma^{-1}$ \cite{archer2002wbg}. We expect the same scenario for the present model fluid.

One important question to address is how robust are the present results. Are the phenomena that we observe simply an artifact of using the simple (RPA) DFT? In to order address this question we have performed further calculations, that we do not describe in detail here, where we have studied the interfacial phase behavior, determined the location of the pre-wetting line and examined the growth of the wetting film using a different (more sophisticated) approximation for the excess Helmholtz free energy functional. The functional that we utilized is the following:
\begin{eqnarray}
F_{ex}[\{\rho_i\}]=F_{ex}[\{\rho_i^b\}] 
+\sum_i \mu_{i}^{ex} \int \dr \delta \rho_i(\rr)\notag \\
-\frac{k_BT}{2}\sum_{i,j} \int \dr \int \dr' \delta \rho_i(\rr)\delta \rho_j(\rr') c_{ij}(|\rr-\rr'|),
\label{eq:F_ex}
\end{eqnarray}
which is obtained by making a Taylor expansion of the excess Helmholtz free energy functional around that of the uniform fluid with densities $\{\rho_i^b\}$ and truncating the expansion at second order in $\delta \rho_i(\rr)\equiv \rho_i(\rr)-\rho_i^b$ \cite{evans1992fif}. $F_{ex}[\{\rho_i^b\}]$ is the excess Helmholtz free energy of the uniform (bulk) system, $\mu_i^{ex}$ is the excess chemical potential of species $i$ in the uniform system and $c_{ij}(r)$ are the pair direct correlation functions in the bulk reference fluid far from the wall. For all of these bulk fluid quantities, that are required as inputs to the theory, we use results obtained from the HNC theory. We find that for the different wall potentials that we considered above within the (RPA) DFT the location of the pre-wetting line obtained from the functional \eqref{eq:F_ex} and the RPA functional \eqref{eq:rpa_func} are located very close together, especially when the moderately small differences in the location of the binodals obtained within the two theories are taken into account \cite{hopkins2006pcf}. Thus we are confident that all of the phenomena that we have observed within the simple RPA theory are at least qualitatively correct. At present we do not know of any computer simulation results for the binary point Yukawa fluid that would confirm this assertion.

Although the point Yukawa model of a fluid is very simplistic, and may be inappropriate for modeling the wide range of real systems where hard-core effects dominate the physics, its usefulness lies in its 
ability to incorporate realistic and relatively complex behavior, without the need for an elaborate density functional theory.
 We have shown that with a very simple model and functional it is possible to describe rich wetting behavior, which is also reproduced in the results of a more sophisticated functional. In order to study effects arising from an explicit hard-core one would require another (hard-sphere) contribution to the functional e.g.~\cite{evans1992fif}, and a significantly greater computational effort to calculate the equilibrium density profiles. This becomes especially important when we move away from simple planar or spherical geometries to situations where the densities vary in two or even three dimensions. For example, we have used the binary point Yukawa model to study the effective interaction between a large colloidal particle and a thick wetting film adsorbed at a planar wall, and a fluid-fluid interface~\cite{hopkins2008pic}. The use of the simple RPA functional greatly decreases the computational cost of calculating individual equilibrium profiles, and of establishing the location and nature of phase transitions.

Interfacial phase behavior similar to that presented here, i.e.\ the existence of a first-order pre-wetting transition and logarithmic growth of thick wetting films has also been observed in other binary systems of soft-core particles namely the Gaussian core and star-polymer solutions~\cite{archer2002bsp,archer2002wbg}. These similarities lead us to conclude that the behavior found for the present model should be quite generic to binary mixtures of purely repulsive particles that exhibit fluid-fluid demixing.

Finally, we note that since the present DFT treatment is purely mean-field, the influence of capillary wave fluctuations in the wetting film interfaces is neglected. In the complete wetting regime fluctuations merely change the amplitude of the logarithmic growth from $\xi$ to $\xi(1+\omega/2)$, where $\omega$ is the standard dimensionless parameter measuring the strength of the fluctuations~\cite{dietrich12pta,archer2002wbg,parry1996tdw}. We believe that a treatment taking fluctuations into account would arrive at the same prediction of a first order pre-wetting phase transition. Where fluctuations are likely to be more important is in the tri-critical regime where the cross-over from first-order to critical wetting occurs.

\begin{acknowledgements}
We gratefully acknowledge stimulating discussions with Matthias Schmidt. PH thanks EPSRC and AJA thanks RCUK for financial support.
\end{acknowledgements}

\appendix

\section{A single order-parameter treatment of the mixture.}
\label{sec:sullivan}
We follow closely the derivation in Refs.\ \cite{sullivan1979vwm,gama1983aaw} for a related model fluid. We begin by setting the external potentials to the exponential type, defined in Eq.~\eqref{eq:exp2_pot_def}.
Substituting Eqs.\ \eqref{eq:yukawa_potential} and \eqref{eq:exp2_pot_def} into the EL equation (\ref{eq:EL_def_1}) gives
\begin{eqnarray}\label{eq:sullivan_1}
\mu_{\mathrm{id},i}(x)&=&\mu_i-A_i \epsilon\exp(-x) \\
& & -\sum_{j=1}^2 \frac{ M_{ij}\epsilon}{2\lambda^{3}}\int_{0}^\infty \dd x'\exp(-|x-x'|)\rho_j(x'), \nonumber
\end{eqnarray}
where $x=\lambda z$ is a dimensionless length. Note that since the density profile is zero for $z<0$, as a consequence of the hard external potential, the lower integral limit becomes $0$. Taking two derivatives with respect to $x$ on both sides of Eq.~\eqref{eq:sullivan_1}
we obtain the following expression:
\begin{equation}
\frac{\dd^2\mu_{\mathrm{id},i}(x)}{\dd x^2}=\mu_{\mathrm{id},i}(x)-\mu_i+\lambda^{-3}\sum_{j=1}^2 M_{ij}\epsilon\rho_j(x).
\label{eq:sullivan_2}
\end{equation}
which is equivalent to that derived in~\cite{gama1983aaw} for a mixture with hard-core pair potentials and Yukawa tails adsorbed at a hard-wall with exponential tails; for the point Yukawa model the hard sphere chemical potential $\mu_{\mathrm{h},i}$ is replaced by the ideal gas chemical potential, $\mu_{\mathrm{id},i}$.

Eq.~\eqref{eq:sullivan_2} constitutes a pair of coupled equations ($i=1,2$). By substituting one into the other and rearranging, it can be shown that these two equations for the fluid density profiles may be made independent of each other if and only if $M_{12}=\sqrt{M_{11}M_{22}}$~\cite{sullivan1982idp}. This corresponds to the ideal geometric mixing rule, where $\delta=0$. It is also necessary for the external potential parameters to be related by the rule: $A_2=\sqrt{M_{22}/M_{11}}A_1$~\cite{gama1983aaw}. Using these parameters the density profiles for the two different species are then related by a simple scaling factor:
\begin{equation}
\mu_{\mathrm{id},2}(x)-\mu_2=\sqrt{\frac{M_{22}}{M_{11}}}(\mu_{\mathrm{id},1}(x)-\mu_1).
\label{eq:sullivan_3}
\end{equation}
Having reduced the problem to determining a single function, it is possible to rearrange Eq.\ \eqref{eq:sullivan_2} to obtain a single ODE for the single order parameter $\mu_{\mathrm{id},i}(x)$ that determines the density profiles of both species: Using the local Gibbs-Duhem relation
\begin{equation}
\rho_i(x)=\frac{\partial p_{\mathrm{id}}(x)}{\partial \mu_{\mathrm{id},i}(x)},\quad i=1,2
\label{eq:duhem}
\end{equation}
where $p_{\mathrm{id}}(x)=k_BT[\rho_1(x)+\rho_2(x)]$ is the local ideal-gas contribution to the pressure and setting $i=1$, we may now write Eq.~\eqref{eq:sullivan_2} as
\begin{equation}
\frac{\dd^2}{\dd x^2}\mu_{\mathrm{id},1}(x)=\mu_{\mathrm{id},1}(x)-\mu_1+M_{11}\epsilon\lambda^{-3}\frac{\dd p_{\mathrm{id}}(x)}{\dd \mu_{\mathrm{id},1}(x)},
\label{eq:sullivan_4}
\end{equation}
where the total derivative is [c.f.\ Eq.~\eqref{eq:sullivan_3}]:
\begin{equation}
\frac{\dd}{\dd \mu_{\mathrm{id},1}(x)}=\frac{\partial}{\partial \mu_{\mathrm{id},1}(x)}+\sqrt{\frac{M_{22}}{M_{11}}}\frac{\partial}{\partial \mu_{\mathrm{id},2}(x)}.
\label{eq:sullivan_5}
\end{equation}
The total pressure, $p$, of the bulk system can be written as
\begin{eqnarray}
\beta p &=& \beta p_{\mathrm{id}}+\frac{\epsilon\lambda^{-3}}{2}[M_{11}(\rho_1^b)^2+2M_{12}\rho_1^b\rho_2^b+M_{22}(\rho_2^b)^2] \nonumber \\
& = & \rho_1^b+\rho_2^b+\frac{M_{11}\epsilon\lambda^{-3}}{2}\left[\rho_1^b+\sqrt{\frac{M_{22}}{M_{11}}}\rho_2^b\right]^2,
\label{eq:sullivan_6}
\end{eqnarray}
where $\rho_i^b$ is the bulk density of species $i$. By integrating both sides of Eq.~\eqref{eq:sullivan_4}, and using the boundary conditions to ensure that the pressure and the chemical potentials tend to their bulk values as $x\to\infty$, we find that
\begin{eqnarray}
\left [\frac{\dd \mu_{\mathrm{id},1}(x)}{\dd x}\right]^2&=&[\mu_{\mathrm{id},1}(x)-\mu_1]^2+2M_{11}\epsilon\lambda^{-3}[p_{\mathrm{id}}(x)-p] \nonumber \\
&\equiv &\psi(\mu_{\mathrm{id},1}).
\label{eq:sullivan_7}
\end{eqnarray}
This equation provides an implicit relation for the chemical potential, $\mu_{\mathrm{id},1}(x)$, and therefore the density profile $\rho_1(x)$. The final step is to integrate:
\begin{equation}
x=\int_{\mu_{\mathrm{id},1}(0)}^{\mu_{\mathrm{id},1}(x)}\frac{\dd \mu_{\mathrm{id},1}}{\pm\sqrt{\psi(\mu_{\mathrm{id},1})}}.
\label{eq:sullivan_8}
\end{equation}

Note that the order parameter $\mu_{\mathrm{id},1}(x)$ which follows is necessarily a monotonic function of $x$ and the choice of sign in \eqref{eq:sullivan_8} depends on whether the order parameter is an increasing or decreasing function of $x$. Since $\mu_{\mathrm{id},1}(x)=\beta^{-1}\ln(\Lambda_1^3\rho_1(x))$ it follows that the density profile $\rho_1(x)$ is also monotonic.
In order to calculate an explicit solution, we must use a numerical method such as the Runge-Kutta method \cite{butcher1987nao}. The only input that is required to this equation is $\mu_{\mathrm{id},1}(x=0)$, the value of the ideal gas chemical potential at the wall. By differentiating Eq.~\eqref{eq:sullivan_1},
evaluating at $x=0$, and substituting back, one finds
\begin{equation}
\left[\frac{\dd \mu_{\mathrm{id},1}(0)}{\dd x}\right]=\mu_{\mathrm{id},1}+2A_1\epsilon-\mu_1.
\label{eq:sullivan_9}
\end{equation}
This may be combined with Eq.~\eqref{eq:sullivan_7}, evaluated at $x=0$, to yield the following relation for $\mu_{\mathrm{id},1}(0)$:
\begin{equation}
 2A_1[\mu_{\mathrm{id},1}(0)+A_1\epsilon-\mu_1]=M_{11}\lambda^{-3}[p_{\mathrm{id}}(0)-p].
\label{eq:sullivan_10}
\end{equation}
Note that if $A_1\to0$ then we recover the contact-density sum-rule for a hard-wall; $\rho_1(0)+\rho_2(0)=\beta p$. Thus, given the particle pair interaction parameters $M_{11}$ and $M_{22}$, and the wall-potential parameter $A_{1}$, one may calculate $\rho_1(x)$ from Eq.\ \eqref{eq:sullivan_8}, where the solution of Eq.~\eqref{eq:sullivan_10} is used as input. One may then calculate $\rho_2(x)$ from Eq.~\eqref{eq:sullivan_3}; $\rho_2(x)$ is also monotonic. Results from this approach are displayed in Sec.~\ref{sec:2cy_wip_results}A.

\end{document}